# Tsallis non - Extensive statistics, intermittent turbulence, SOC and chaos in the solar plasma.
# Part two: Solar Flares dynamics

L.P. Karakatsanis[1], G.P. Pavlos[1] M.N. Xenakis[2]

*[1] Department of Electrical and Computer Engineering, Democritus University of Thrace, 67100 Xanthi, Greece*
*[2] German Research School for Simulation Sciences, Aachen, Germany*
*Email : karaka@xan.duth.gr, lkaraka@gmail.com*

## Abstract

*In the second part of this study and similarly with part one, the nonlinear analysis of the solar flares index is embedded in the non-extensive statistical theory of Tsallis [1]. The $q-triplet$ of Tsallis, as well as the correlation dimension and the Lyapunov exponent spectrum were estimated for the SVD components of the solar flares timeseries. Also the multifractal scaling exponent spectrum $f(a)$, the generalized Renyi dimension spectrum $D(q)$ and the spectrum $J(p)$ of the structure function exponents were estimated experimentally and theoretically by using the $q-entropy$ principle included in Tsallis non extensive statistical theory, following Arimitsu and Arimitsu [2]. Our analysis showed clearly the following: a) a phase transition process in the solar flare dynamics from high dimensional non Gaussian SOC state to a low dimensional also non Gaussian chaotic state, b) strong intermittent solar corona turbulence and anomalous (multifractal) diffusion solar corona process, which is strengthened as the solar corona dynamics makes phase transition to low dimensional chaos: c) faithful agreement of Tsallis non equilibrium statistical theory with the experimental estimations of i) non-Gaussian probability distribution function $P(x)$, ii) multifractal scaling exponent spectrum $f(a)$ and generalized Renyi dimension spectrum $D_q$, iii) exponent spectrum $J(p)$ of the structure functions estimated for the sunspot index and its underlying non equilibrium solar dynamics. e) The solar flare dynamical profile is revealed similar to the dynamical profile of the solar convection zone as far as the phase transition process from SOC to chaos state. However the solar low corona (solar flare) dynamical characteristics can be clearly discriminated from the dynamical characteristics of the solar convection zone.*

## 1. Introduction

In the first part of this study [3] we have introduced a new theoretical framework for studying solar plasma dynamics by using experimental time series observations. The novelty of this framework includes the new concepts of Tsallis q-statistics theory coupled with intermittent turbulence theory [8] and chaos theory. These theoretical concepts were used for the extension of an existed chaotic algorithm for experimental time series analysis [3]. In part one the sunspot dynamics was studied from the new point of view of Tsallis q-statistics of solar dynamics coupled with intermittent turbulence of the convective zone solar plasma and the self-organized criticality (SOC) and low dimensional perspectives. In this part two of our study we extend the analysis of part one to the solar flare plasma dynamics.

As the Tsallis q-statistics extension of Boltzmann-Gibbs statistics is harmonized with the fractal generalization of dynamics [1,4,5], our extended algorithm of time series analysis revealed new possibilities for the verification of significant theoretical concepts. Especially it is of high theoretical interest the experimental verification of chaotic multiscale and multifractal dynamics underlying the experimental time series observations. As we show in this paper the coupling of q-statistics with fractal dynamics and low dimensional chaos introduces new and significant knowledge about the solar flare dynamics.

Moreover the solar flare dynamical system is of high interest as at solar flare regions can exist strong turbulent magnetic energy dissipation and strong acceleration charged particles as electrons, protons and heavy ions. These bursty phenomena are followed by hard X-ray luminosity, while the underlying mechanism of solar flare dynamics is some kind of magnetic reconnection process. However, the proposed known physical explanations of the classical magnetic reconnection process are inevitable to explain the structure dynamics of magnetic field dissipation [6,7]. In this study, we indicate the existence of fractal dissipation – acceleration mechanism in accordance to Pavlos [8].

In the following of this study in section (2) we present the presupposed theoretical tools for the analysis of solar flare index time series. In section (3) we present the results of the data analysis. In section (4) we summarize the data analysis results, while in section (5) we discuss our results.

## 2. Theoretical Concepts and Methodology of Data Analysis

In the section two of the first part of this study [3] we have presented the theoretical pre-suppositions for the data analysis. In these pre-suppositions we have included analytically the following:

a) The physical meaning of Tsallis non-extensive entropy theory (Part one, sub sections 2.11-2.13). It is significant to memorize that far from equilibrium it is possible to be developed spatiotemporal plasma structures including long-range correlations. This can be indicated by the estimation of Tsallis q-triplet. Especially, far from equilibrium the q-triplet ($q_{sen}, q_{stat}, q_{rel}$) can differ significantly from the equilibrium Gaussian profile where $q_{sen} = q_{stat} = q_{rel} = 1$.

b) Tsallis non-extensive entropy theory is related with the multi-fractal and multi-scale character of the underlying phase space which far from equilibrium includes anomalous topology related with multi-scaling and multi-fractality. These characters are related with corresponding multifractal and anomalous topology of the dissipation regions in the physical space-time. For this we estimate the multifractal spectrum of dimensions $f(a)$ and $D_q$ according to multi-fractal theory as well as the structure function ($S_p$) and its scaling exponent spectrum [$J(p)$] which can differ from the Kolmogorov's first theory (K41) predictions. According to Frisch [9] at the non-Gaussian multi-fractal and intermittent dissipation processes, the scaling exponent spectrum $J(p)$ satisfies the relation:

$$\frac{dJ(p)}{dp} = h_*(p) \neq 0 \tag{1}$$

Where $h_*(p)$ is related to the fractal dimension $D(h)$ of the fractal dissipation region by the relation:

$$D'(h_*(p)) = p \tag{2}$$

The fractal dimension $D(h)$ and the exponents of the structure function $J(p)$ are related with a Legendre transformation:

$$J(p) = \inf_{(h)} \left[ ph + 3 - D(h) \right] \tag{3}$$

This relation indicates the fact that, when the dissipation region is a multi-fractal then $D(h) \neq 3$ and $dJ(p)/dp \neq p$, while for k41 theory of Kolmogorov the dissipation region is mono-fractal.

c) The correlation dimensions is given by the relation $D = \lim_{\substack{r \to 0 \\ m \to \infty}} D_m$, where $D_m = \lim_{r \to 0} \frac{d \ln C_m(\mathbf{r}, m)}{d \ln(r)}$ and $C_m(r, m) = \frac{2}{N(N-1)} \sum_{i=1}^{N} \sum_{j=1}^{N} \Theta(r - \|x(i) - x(j)\|)$. For the estimation of $D$ we follow the theories of Takens [10], Grassberger and Procaccia [11] and Theiler [12].

d) The Lyapunov exponent spectrum ($\lambda_i$) obtained by the perturbations of the dynamical in the reconstructed space [13].

e) Discrimination of colored noise of pseudo chaos profile from the low dimensional chaotic (random) data by using the Theiler method of Surrogate data [14].

According to the above theoretical concepts and the analytic description of section (2) of the first part of this study [3] we follow the next plan for the experimental data analysis of the solar flare index:

I) We use the Singular Value Decomposition for the for the discrimination of deterministic and noisy (stochastic) components included in the observed signals, as well as for the discrimination of distinct dynamical components

II) Estimation of the Flatness coefficient F which for values F>3.

III) Estimation of the $q_{stat}$ index of the q-Gaussian through linear correlation fitting of $\ln_q p(z)$ versus $z^2$, where $\ln_q p(z)$ is the q-logarithmic function of the probability function $p(z)$, as $z = z_{n+1} - z_n$, ($n = 1, ..., N$) corresponds to the first difference of the experimental solar flares time series data [15].

IV) Determination of the $q_{rel}$ index of the q-statistics according to the relation: $\frac{d\Omega}{dt} = -\frac{1}{T_{q_{rel}}} \Omega^{q_{rel}}$, where corresponds to the autocorrelation functions or the mutual information function of the experimental time series [1]

V) Determination of the $q_{sen}$ index according to the relation: $\frac{1}{q_{sen}} = \frac{1}{a_{min}} - \frac{1}{a_{max}}$, where corresponds to the zero points of the multifractal exponent spectrum. The index is related to the Kolmogorov-signal entropy production and Pesin theory [16,17].

VI) Determination a) of the structure functions $S(R) = \langle |\delta u|^p \rangle$, where $\delta u$ is the spatial variation of the bulk plasma flow velocity and b) of the scaling exponent spectrum $J(p)$ according to the relation: $S(p) \sim l^{J(p)}$, where $l$ is the length scale of dissipation process [8]

VII) Determination of the correlation dimension (D) by using the saturation value of the slopes (Dm) of the correlation integrals (Cm)

VIII) Determination of the Lyapunov exponent's spectrum.

IX) Determination of the significance (σ) of the discriminating statistics by using the surrogate method of Theiler [14]. For σ-values >3 the null hypothesis can be rejected with confidence higher than 99%.

## 3. Results of Data Analysis

In this section we analyse the daily Flare Index of the solar activity that was determined using the final grouped solar flares obtained by NGDC (National Geophysical Data Center). It is calculated for each flare using the formula: $Q = (i*t)$, where "$i$" is the importance coefficient of the flare and "$t$" is the duration of the flare in minutes. To obtain final daily values, the daily sums of the index for the total surface are divided by the total time of observation of that day. The data covers time period from 1/1/1996 to 31/12/2007.
In this section we present results concerning the analysis of data included in the sunspot index by following physical and the methodology included in the previous section of this study.

### 3.1 Time series and Flatness Coefficient F.

Fig.1a represents the Solar Flares time series that was constructed concerning the period of 11 years. Fig1b presents the flatness coefficient F estimated for the solar flare data during the same period. The F values reveal continues variation of the flares statistics between Gaussian profile ($F \sim 3$) to strong non-Gaussian profile ($F \sim 4-25$). Fig.1c presents the first ($V_1$) SVD component of the solar flare index and Fig.1e presents the sum $V_{2-15} = \sum_{i=2}^{15} V_i$ of the next SVD components estimated for the sunspot index time series. The estimation of the flatness coefficient $F(V_1)$ and $F(V_{2-15})$ is shown in Fig.1d and Fig.1f, correspondingly. As we can notice in these figures the statistics of the $V_1$ component is clearly discriminated from the statistics of the $V_{2-15}$ component, as the $F(V_1)$ flatness coefficient obtained almost everywhere low values ($\sim 3-5$) while the $F(V_{2-15})$ obtained much higher values($\sim 5-30$). The low values ($\sim 3-4$) of the $F(V_1)$ coefficient indicates for the solar activity a near Gaussian dynamical process, underlying the $V_1$ SVD component. Oppositely, the high values of the $F(V_{2-15})$ coefficient indicate strongly non-Gaussian solar corona dynamical process underlying the $V_{2-15}$ SVD component.

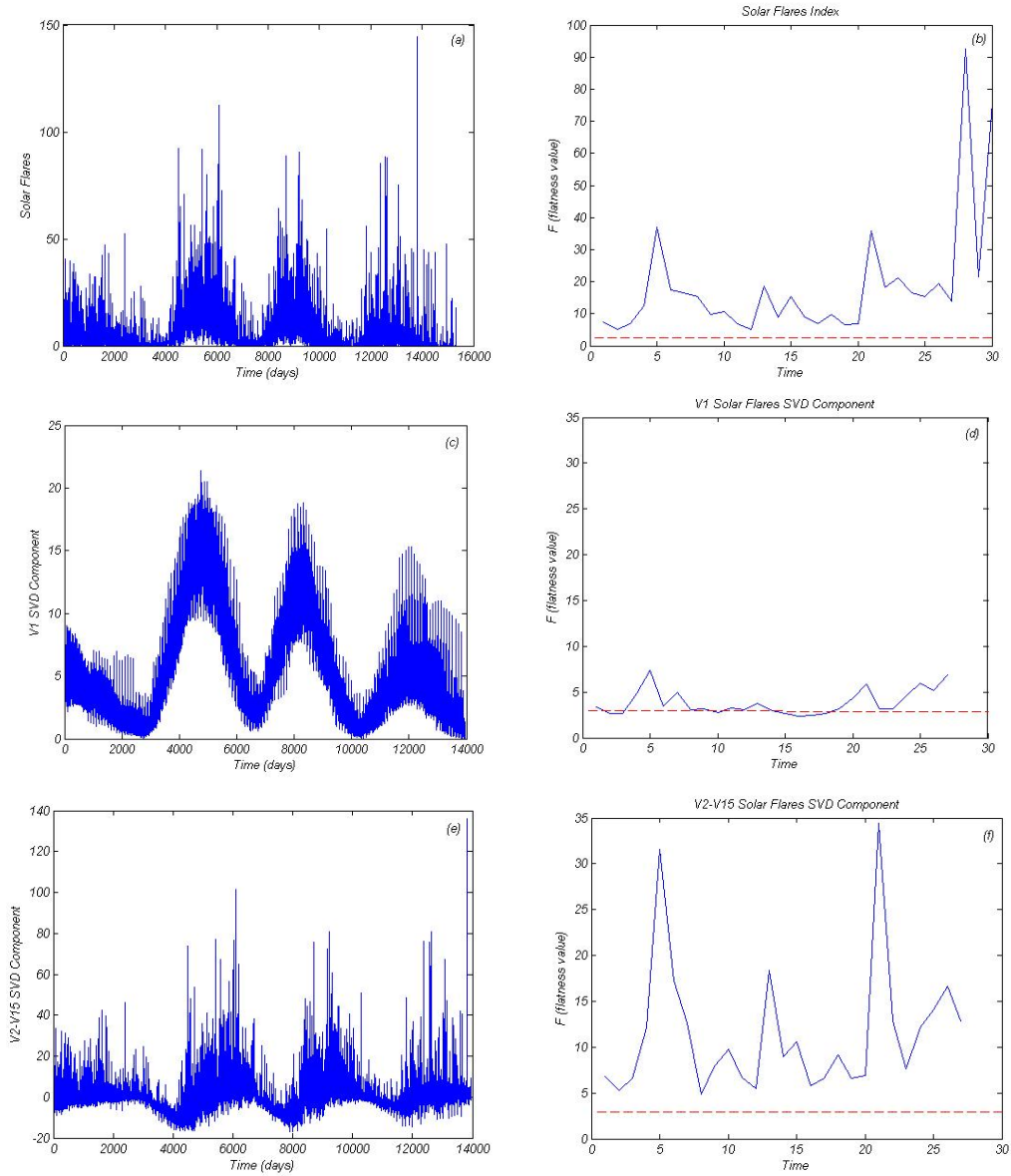

***Figure 1****: (**a**) Time series of Solar Flares Index concerning the period of 11 years. (**b**) The coefficient F estimated for Solar Flares Index TMS (**c**) The SVD $V_1$ component of Solar Flares Index tms. (**d**) The coefficient F estimated for the SVD $V_1$ component of Solar Flares Index tms. (**e**) The SVD $V_{2-15}$ component of Solar Flares Index tms. (**f**) The coefficient F estimated for the SVD $V_{2-15}$ component of Solar Flares Index tms.*

## 3.2 The Tsallis q-statistics

In this section we present results concerning the computation of the Tsallis q-triplet, including the three-index set ($q_{stat}, q_{sen}, q_{rel}$) estimated for the original solar flare index timeseries, as well as for its $V_1$ and $V_{2-15}$ SVD components (presented in Fig.(2-4)).

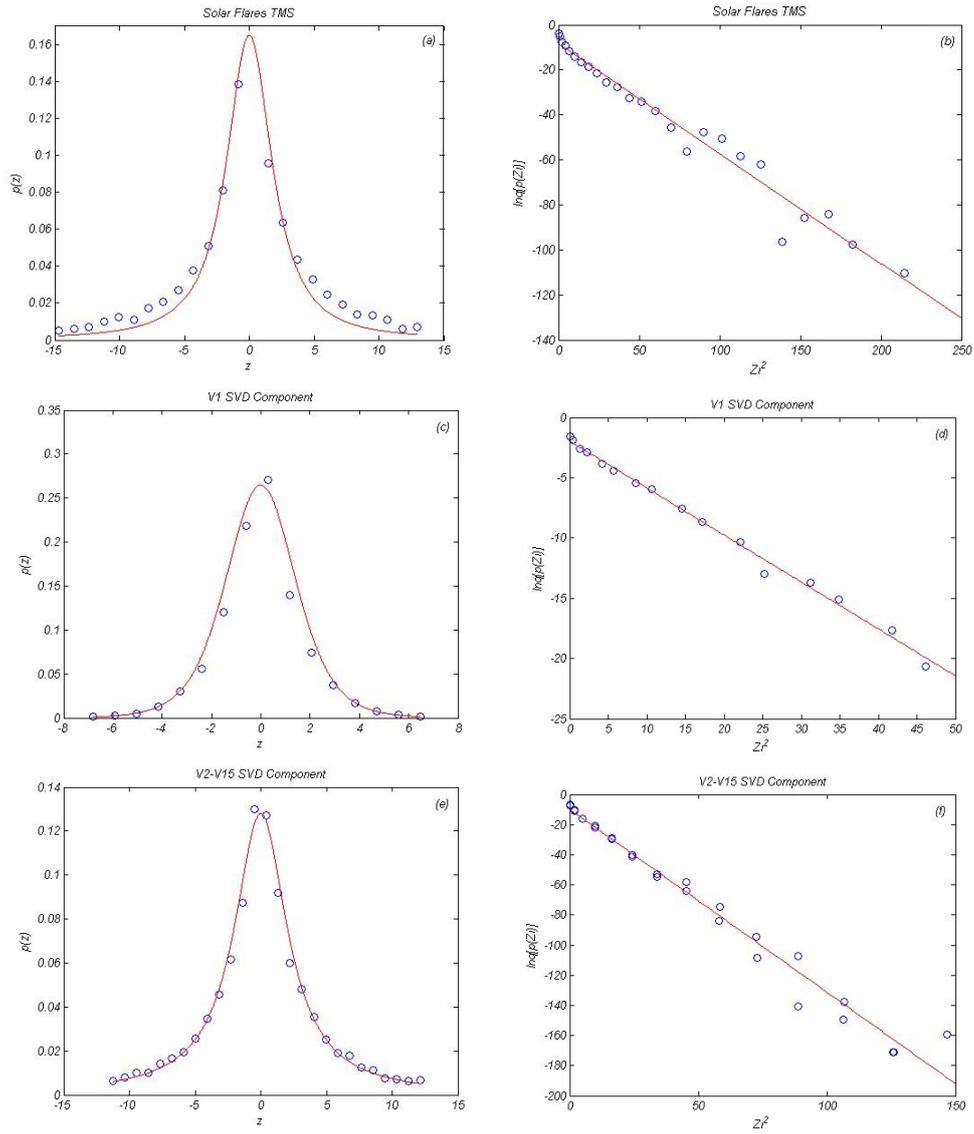

***Figure 2:*** *(**a**) PDF P($z_i$) vs. $z_i$ q Guassian function that fits P($z_i$) for the Solar Flares Index timeseries (**b**) Linear Correlation between $\ln_q p(z_i)$ and $(z_i)^2$ where q = 1.87±0.06 for the Solar Flares Index timeseries (**c**) PDF P($z_i$) vs. $z_i$ q Gaussian function that fits P($z_i$) for the $V_1$ SVD component (**d**) Linear Correlation between $\ln_q p(z_i)$ and $(z_i)^2$ where q = 1.28±0.04 for the $V_1$ SVD component (**e**) PDF P($z_i$) vs. $z_i$ q Gaussian function that fits P($z_i$) for the $V_{2-15}$ SVD component (**f**) Linear Correlation between $\ln_q p(z_i)$ and $(z_i)^2$ where q = 2.02±0.15 for the $V_{2-15}$ SVD component*

### 3.2.1 Determination of $q_{stat}$ index of the q-statistics.

In Fig.2a we present (by open circles) the experimental probability distribution function (PDF) p(z) vs. z, where z corresponds to the $Z_{n+1} - Z_n, (n = 1, 2, ..., N)$ timeseries difference values. In Fig.2b we present the best linear correlation between $\ln_q[p(z)]$ and $z^2$. The best fitting was found for the value of $q_{stat} = 1.87 \pm 0.06$. This value was used to estimate the q-Gaussian distribution presented in Fig.2a by the solid black line. Fig.2[c,d] and Fig.2[e,f] are similar to Fig.2[a,b] but

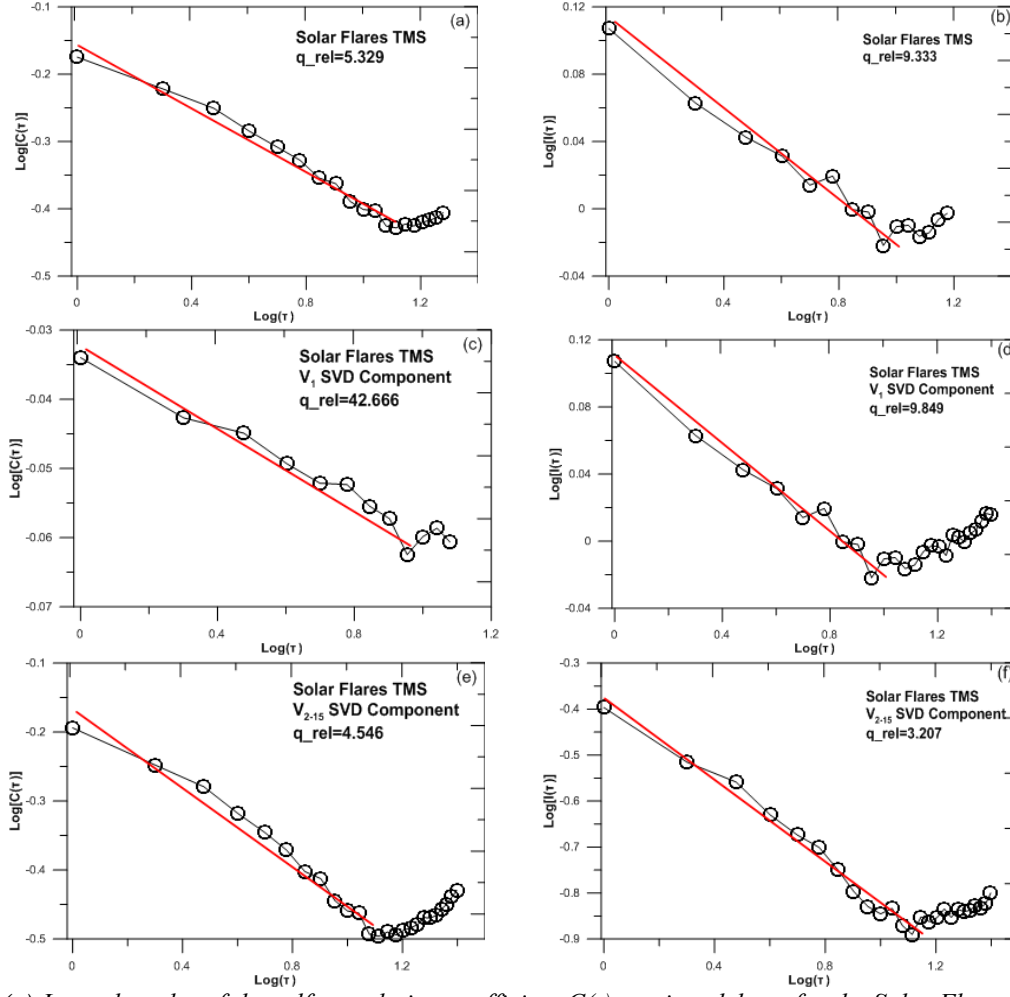

*Figure 3*: *(a) Log – log plot of the self-correlation coefficient C(τ) vs. time delay τ for the Solar Flarest Index time series. We obtain the best fit with qrel=5.329±0.135 (b)Log – log plot of the mutual information I(τ) vs. time delay τ. for the Sunspot Index time series. We obtain the best fit with qrel=9.333±0.395 (c) Log – log plot of the self-correlation coefficient C(τ) vs. time delay τ for the $V_1$ SVD component. We obtain the best fit with qrel=42.666±2.950 (d)Log – log plot of the mutual information I(τ) vs. time delay τ for the $V_1$ SVD component. We obtain the best fit with qrel=9.849±0.410. (e) Log – log plot of the self-correlation coefficient C(τ) vs. time delay τ for the $V_{2-15}$ SVD component. We obtain the best fit with qrel=4.546±0.173 (f)Log – log plot of the mutual information I(τ) vs. time delay τ. for the $V_{2-15}$ SVD component. We obtain the best fit with qrel=3.207±0.086.*

for the $V_1$ and $V_{2-15}$ SVD components correspondingly. Now as concerns the SVD components, the $q_{stat}$ values were found to be: $q_{stat}(V_1) = 1.28 \pm 0.04$ and $q_{stat}(V_{2-15}) = 2.02 \pm 0.15$. As we can observe from these results the following relation is satisfied:

$1 < q_{stat}(V_1) < q_{stat}(orig) < q_{stat}(V_{2-15})$, where $q_{stat}(orig)$ corresponds to the original solar flare index timeseries.

### 3.2.2 Determination of $q_{rel}$ index of the q-statistics.
#### a) Relaxation of autocorrelation functions

Fig.3 presents the best log plot fitting of the autocorrelation function $C(\tau)$ estimated for the original solar flare index signal (Fig.3a) its $V_1$ SVD component (Fig.3c), as well as its $V_{2-15}$ SVD component (Fig.3e). The three $q_{rel}$ values were found to satisfy the relation: $1 < q_{rel}^c(V_{2-15}) < q_{rel}^c(orig) < q_{rel}^c(V_1)$ as:
$q_{rel}^c(V_{2-10}) = 4.546 \pm 0.173$, $q_{rel}^c(V_{orig}) = 5.329 \pm 0.135$, $q_{rel}^c(V_1) = 42.666 \pm 2.950$

**b) Relaxation of Mutual Information**

Fig.3[b,d,f] is similar with Fig.3[a,c,e] but it corresponds to the relaxation time of the mutual information $I(\tau)$. For the top of the bottom we see the log-log plot of $I(\tau)$ for the solar flare index timeseries, its $V_1$ SVD component and its $V_{2-15}$ SVD component. The best log-log (linear) fitting showed the values:
$q_{rel}^I(V_{2-10}) = 3.207 \pm 0.086$, $q_{rel}^I(V_{orig}) = 9.333 \pm 0.395$, $q_{rel}^I(V_1) = 9.849 \pm 0.410$. Among the three values it is satisfied the following relation: $1 < q_{rel}^I(V_{2-10}) < q_{rel}^I(orig) < q_{rel}^I(V_1)$. Also comparing the $q_{rel}$ indices, as they were estimated for the autocorrelation function and the mutual information function, it was found that the values are different for all the cases. The last result is explained by the fact that the mutual information includes nonlinear characteristics of the underlying dynamics in contrast to the autocorrelation function which is a linear statistical index.

### 3.2.3 Determination of $q_{sen}$ index of the q-statistics.

Fig.4 presents the estimation of the generalized dimension $D_q$ and their corresponding multifractal (or singularity) spectrum $f(\alpha)$. The $q_{sen}$ index was estimated by using the relation $1/(1-q_{sens}) = 1/a_{min} - 1/a_{max}$ for the original solar flares index (Fig.4 [a-b], as well as for its $V_1$ (Fig.4[c-d]) and $V_{2-10}$ (Fig.4[e-f]) SVD components. The three index $q_{sen}$ values were found to satisfy the relation: $1 < q_{sen}(V_1) < q_{sen}(V_{2-10}) < q_{sen}(orig)$, where: $q_{sen}(V_1) = -0.540$, $q_{sen}(V_{2-15}) = 0.192$, and $q_{sen}(V^{orig}) = 0.308$.

In Fig.4[a,c,e] the experimentally estimated spectrum function $f(\alpha)$ is compared with a polynomial of sixth order (solid line) as well as by the theoretically estimated function $f(\alpha)$ (dashed line), by using the Tsallis $q$–entropy principle and following Arimitsu and Arimitsu [2]. As we can observe the theoretical estimation is faithful with high precision on the left part of the experimental function $f(a)$. However, the fit of theoretical and experimental data are less faithful for the right part of $f(a)$, especially for the original solar flares timeseries (Fig.4a) and its SVD components $V_1$ (Fig.4c). Finally, the coincidence of theoretically and experimentally data is excellent for the $V_{2-15}$ SVD component (Fig.4e).

Similar comparison of the theoretical prediction and the experimental estimation of the generalized dimensions function $D(q)$ is shown in Fig.4(b,d,f). In these figures the solid brown line correspond to the $p$–model prediction according to [18], while the solid red line correspond to the $D(q)$ function estimation according to Tsallis theory [1].

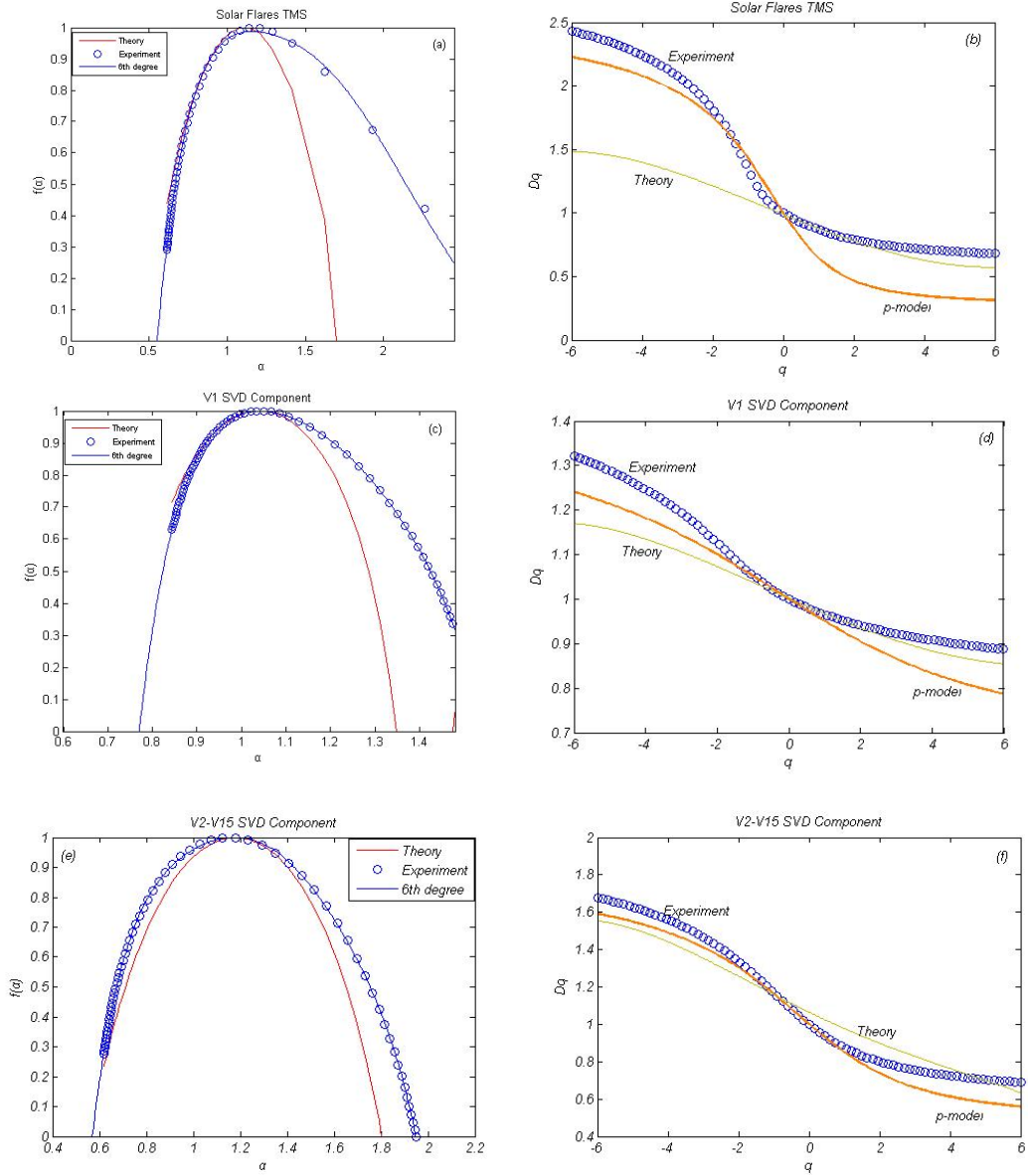

*Fig4: (a) Multifractal spectrum of Solar Flares timeseries. With solid line a sixth degree polynomial. We calculate the q_sen=0.308. (b) D(q) vs. q of Sunspot timeseries. (c) Multifractal spectrum of $V_1$ SVD component. We calculate the q_sen=-0.540. With solid line a sixth degree polynomial (d) D(q) vs. q of $V_1$ SVD component. (e) Multifractal spectrum of $V_{2-15}$ SVD component. We calculate the q_sen=0.192 . With solid line a sixth degree polynomial (f) D(q) vs. q of $V_{2-15}$ SVD component.*

The correlation coefficient of the fitting was found higher than 0.9 for all cases. These results indicate for the turbulence cascade of solar corona plasma, partial mixing and asymmetric (intermittent) fragmentation process of the energy dissipation.

It is interesting here to notice also the relation between the $\Delta\alpha_{\min,\max}$ values where $\Delta\alpha_{\min,\max} = a_{\max} - \alpha_{\min}$, which was found to satisfy the following ordering relation:

$$\Delta\alpha(V_1) = 0.77 < \Delta\alpha(V_{2-10}) = 1.37 < \Delta\alpha(orig) = 2.15,$$

The $\Delta D_{-\infty,+\infty}$ and $\Delta \alpha_{min,max}$ values were estimated also separately for the first fifteen SVD components. The $\Delta D_{-\infty,+\infty}(V_i)$ values vs. $V_i, (i=1,...,15)$ are showed in Fig.5a and the $\Delta \alpha_{min,max}(V_i)$ vs. $V_i, (i=1,...,15)$ are showed in Fig.5b. The spectra of $\Delta D_q$ and $\Delta D_a$ shown in these figures (Fig.5[a,b]) reveal positive and increasing profile as we pass from the first to the last SVD component. This clearly indicated intermittent solar turbulent process underlying all the SVD components of the solar flares index. However the intermittency character becomes stronger at large SVD components.

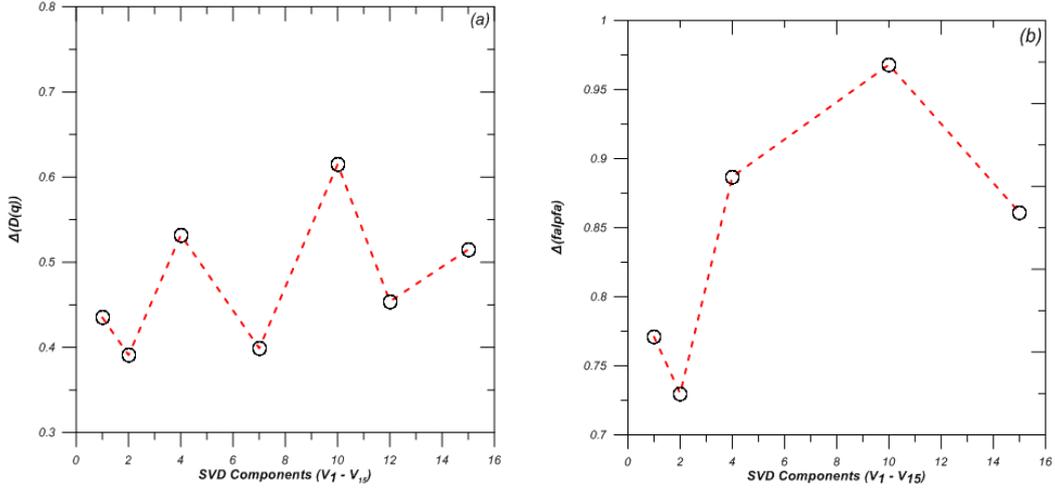

*Figure 5: (a) The differences Δ(Dq(Vi)) versus the Vi SVD components of Solar Flarest timeseries for i=1,2,...,15. (b) The differences Δ(Da(Vi)) versus the Vi SVD components of Solar flares timeseries for i=1,2,...,15.*

### 3.3 Determination of Structure Function spectrum
### 3.3.1 Intermittent Solar Turbulence

Fig.6[a-c-e] shows the structure function $S(p)$ plotted versus time lag ($\tau$) estimated for the original solar flare index signal (Fig.6a) as well as for its SVD components $V_1$ (Fig.6c) and $V_{2-15}$ (Fig.6e). At low values of time lag ($\tau$), we can observe scaling profile for all the cases of the original timeseries and its SVD components for every p value. Fig.6[b-d-f] presents the best linear fitting of the scaling regions in the lag time interval $\Delta \log(\tau) = 0 - 1.4$. Fig.7a shows the exponent $J(p)$ spectrum of structure function vs. the $p$th order estimated separately for the first seven SVD components $V_i, (i=1,...,15)$. In the same figure we present the exponent $J(p)$ spectrum of the structure function $S(p)$ according to the original theory of Kolmogorov [68] for the fully developed turbulence, of Gaussian turbulence known as $p/3$ theory. Here we can observe clear discrimination from the Gaussian turbulence K41 theory for all the SVD components. Moreover, we observe significant dispersion of the $J(p)$ spectrum from the first to the fifteenth SVD component, especially for the higher orders ($p > 10$).

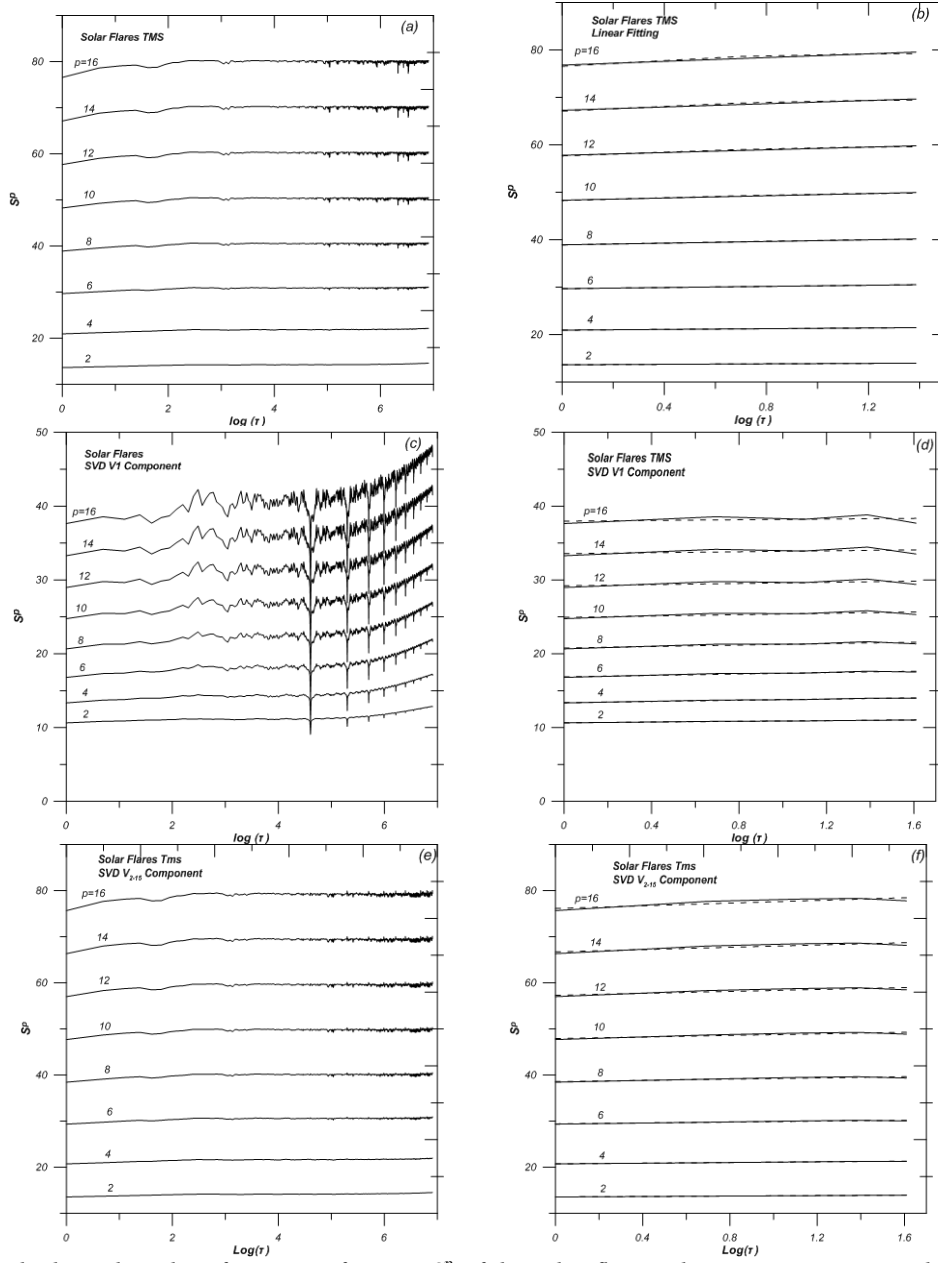

*Figure 6: (a) The log – log plot of structure function $S^p$ of the solar flare index timeseries vs. time lag τ for various values of the order parameter p. (b) The first linear scaling of the log – log plot. (c) The log – log plot of structure function $S^p$ of the $V_1$ SVD component vs. time lag τ for various values of the order parameter p. (d) The first linear scaling of the log – log plot. (e) The log – log plot of structure function $S^p$ of the $V_{2-15}$ SVD component vs. time lag τ for various values of the order parameter p. (f) The first linear scaling of the log – log plot.*

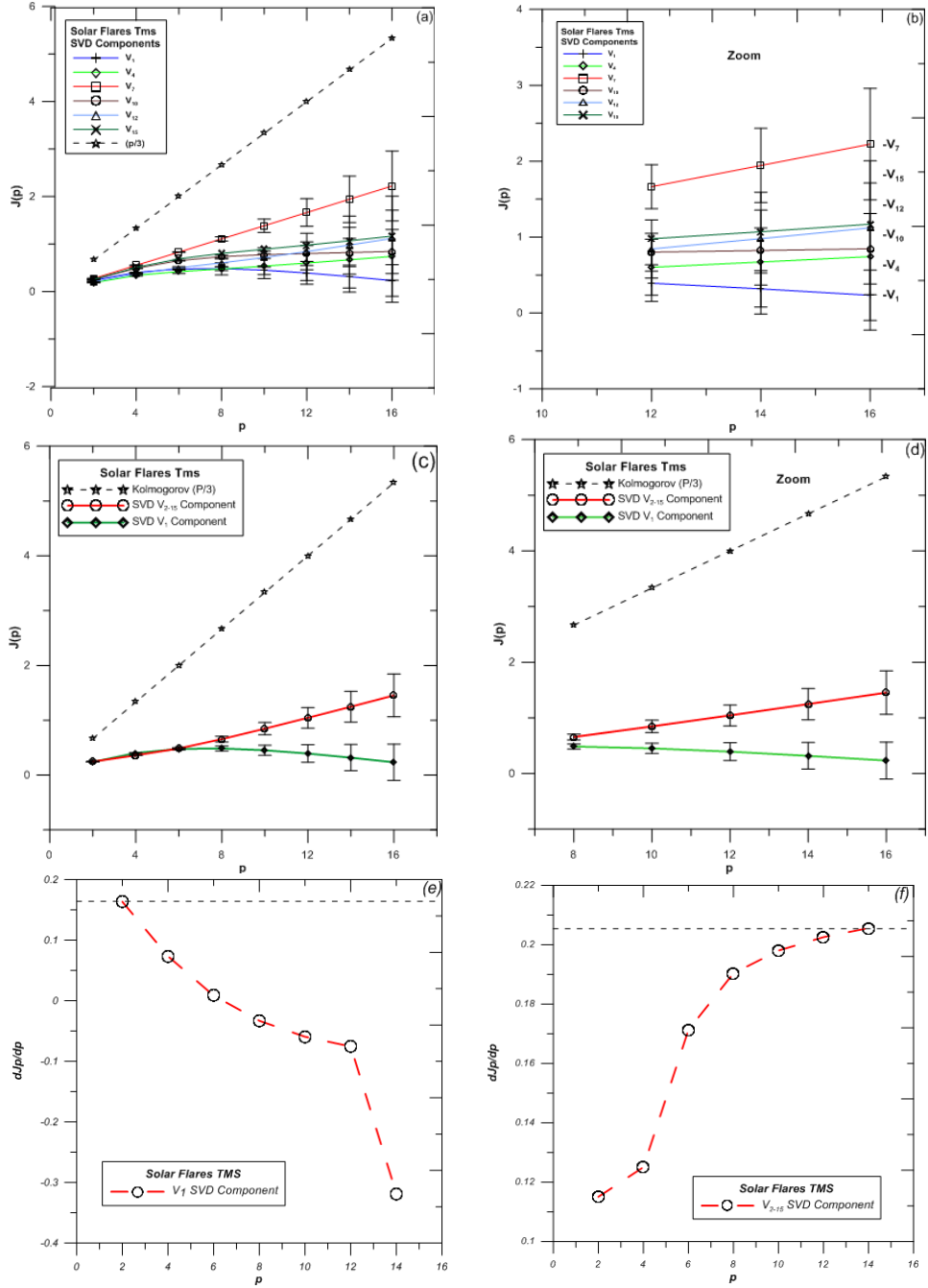

*Figure 7: (a) The scaling exponent J(p) versus p of the independent $V_1$-$V_{15}$ SVD components of the Solar Flares timeseries and compared with the Kolmogorov p/3 prediction (dashed line) (b) The zoom in the area of p=12-16. The scaling exponent J(p) versus p of the $V_1$, $V_{2-15}$ SVD components of the solar flares timeseries and compared with the Kolmogorov p/3 prediction (dashed line) (b) The zoom in the area of p=8-16. (e)The h(p)=dJ(p)/dp function versus p for the $V_1$ SVD component of Solar Flares timeseries. (f)The h(p)=dJ(p)/dp function versus p for the $V_{2-15}$ SVD component of Solar Flares timeseries.*

Fig.7c is similar to Fig.7a but for the $V_1$ SVD component and the summarized $V_{2-15}$ SVD component. Also, here we notice significant discrimination between the $V_1$ and $V_{2-15}$ SVD components while both of them reveal *pth* order structure functions with their own index different from the $p/3$ values given by the K41 theory. In order to study furthermore the departure from the Gaussian behavior of the turbulence underlying the solar flares index signal, we present in Fig.7[e-f] the derivatives $h(p) = dJ(p)/dp$ of the structure functions estimated for the spectra $S(p)$ of the $V_1$ and $V_{2-10}$ SVD components. For both cases the non-linearity of the functions $S_{V_1}(p), S_{V_{2-10}}(p)$ is apparent as their derivatives are strongly dependent upon the $p$ values.

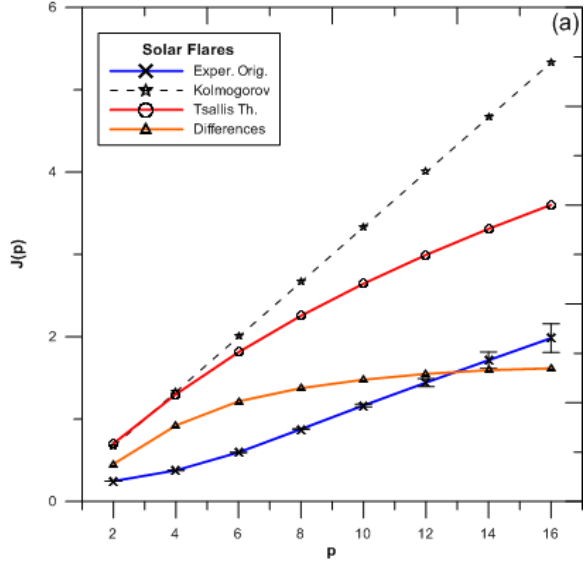
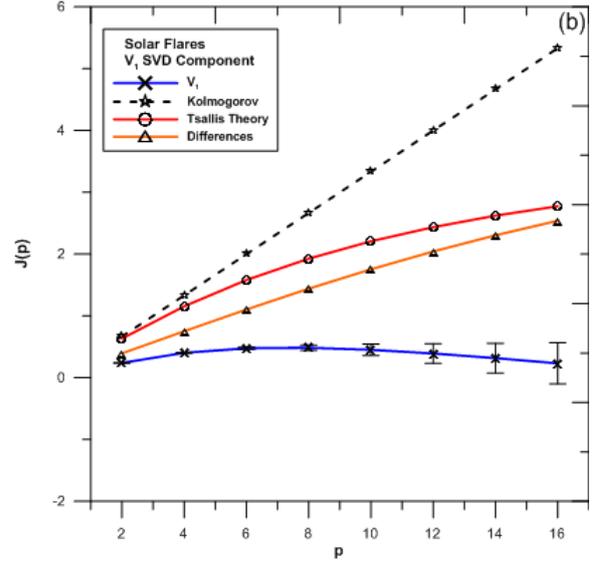
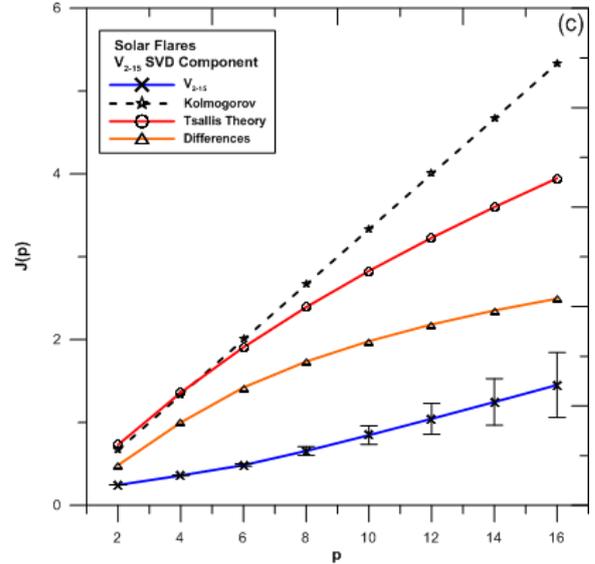

*Figure 8: (a) The J(p) function versus p for sunspot timeseries and compared with the Kolmogorov (p/3) prediction (dashed line), the theoretical curve of Tsallis theory and its differences between experimental and theoretical. (b) The J(p) function versus p for $V_1$ SVD component compared with the Kolmogorov (p/3) prediction (dashed line), the theoretical curve of Tsallis theory and its differences between experimental and theoretical results. (c) The J(p) function versus p for $V_{2-10}$ SVD component and compared with the Kolmogorov (p/3) prediction (dashed line), the theoretical curve of Tsallis theory and its differences between experimental and theoretical results.*

### 3.3.2` Comparison of solar turbulence with non-extensive q-statistics.

In this section we present interesting results concerning the comparison of the structure function experimental estimation with the theoretical predictions according to Arimitsu and Arimitsu [2] by using the Tsallis non-extensive statistical theory, as it was presented at the description of section 2 of part one of this study [3]. Fig.8a presents the structure function $J(p)$ vs. the order parameter $p$, estimated for the solar flares index timeseries (black line) the K41 theory (dashed line), the theoretically predicted values of structure function (red line) by using q-statistics theory, as well as the differences $\Delta J(p)$ vs. $p$, between the experimentally and theoretically produced values. Fig.8[b-c] is similar to Fig.8a but corresponding to the $V_1$ and $V_{2-15}$ SVD components of the original solar flares index timeseries. For all the cases presented in Fig.8[a-c] the theoretically estimated $J(p)$ values (red lines) correspond to the HD turbulent dissipation of the solar plasma, revealing values lower than the K41 prediction in accordance with the HD intermittent turbulence.

According to previous theoretical description (section 2) of part one the experimentally produced structure function spectrum of the original signal and its SVD components is caused by including kinetic and magnetic dissipation simultaneously [7] of the MHD solar turbulence. In this case the solar corona magnetic field dissipation makes the structure function spectrum to obtain values much different than the values of the corresponding HD intermittent turbulence. As we can observe in Fig.8 [a-b] the best fitting to the differences $\Delta J(p)$ shows for all cases (the sunspot index and its $V_1$, $V_{2-15}$ SVD components) the existence of the linear relation: $\Delta J(p) \approx \alpha(p+b)$ at high values of $p$.

## 3.5 Determination of Correlation Dimension

### 3.5.1 Correlation dimension of the Sunspot timeseries

Fig.9[a] shows the slopes of the correlation integrals vs. $\log r$ estimated for the solar flares index timeseries for embedding dimension $m = 6-10$. As we notice here there is no tendency for low value saturation of the slopes which increase continuously as the embedding dimension ($m$) increases. The comparison of the slopes with surrogate data, showed in Fig.9[b-c]. Fig.9b presents the slopes of the original signal at embedding dimension $m = 7$ (red line), as well as the slopes estimated for a group of corresponding surrogate data. The significance of the statistics was shown in Fig.9[d]. As the significance of the discriminating statistics remains much lower than two sigmas, we cannot reject the null hypothesis of a high dimensional Gaussian and linear dynamics underlying the Sunspot index of the solar activity.

Fig.10 shows results concerning the estimation correlations dimensions of the $V_1$ and $V_{2-15}$ SVD components of the original sunspot index signal.

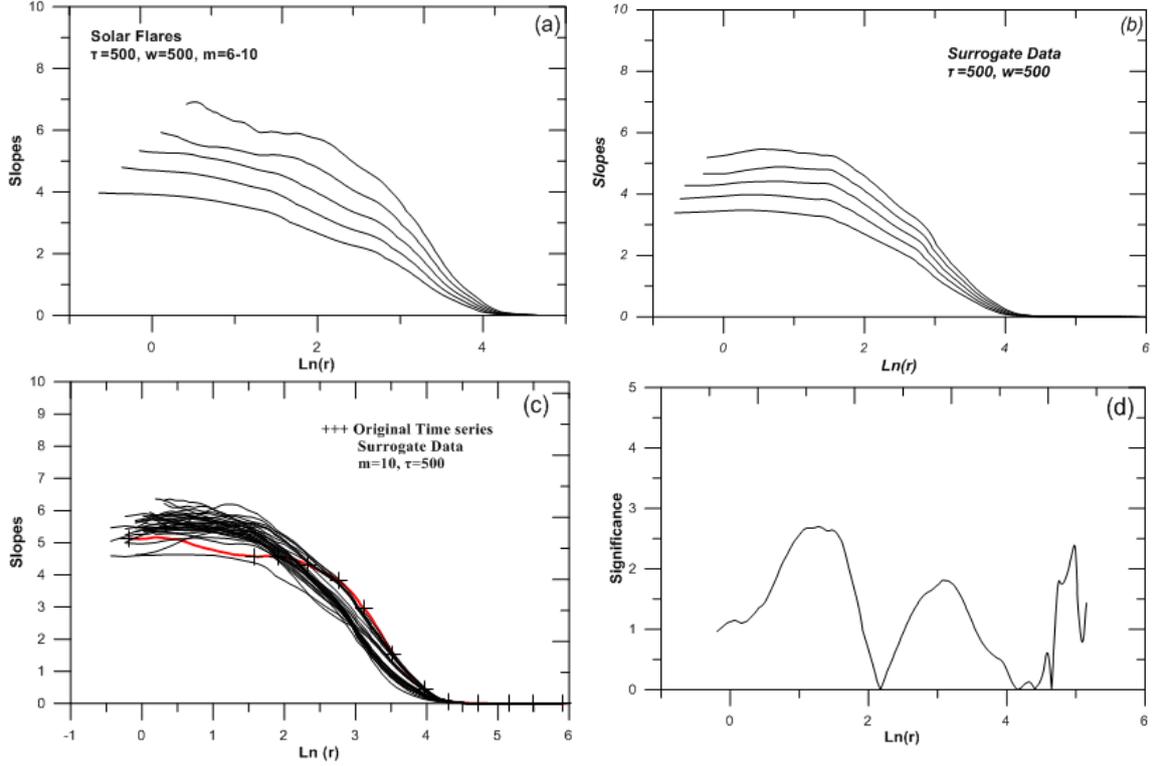

***Figure 9:*** *(**a**) Slopes D of the correlation integrals estimated for the Solar Flares timeseries. (**b**) Slopes D of the correlation integrals estimated for the Surrogate timeseries. (**c**) Slopes of the correlation integrals of the solar flares time series and its thirty (30) surrogate estimated for delay time τ=500 and for embedding dimension m=10, as a function of Ln(r). (**d**) The Significance of the Statistics for the solar flares timeseries and its 30 Surrogates.*

### 3.5.1 Correlation dimension of the $V_1$ SVD component

The slopes of the correlation integrals estimated for the $V_1$ SVD component and its corresponding surrogate data are presented in Fig.10a and Fig.10b. The slope profiles of the $V_1$ SVD component reveal no saturation. However, the corresponding slope profiles of the surrogate data are similar with the slopes profile of the $V_1$ SVD component. Fig.10c shows the slopes of the $V_1$ SVD component and a group of corresponding surrogate data at the embedding dimension $m = 10$, while in Fig.10d we present the significance of the statistics. As the significance remains at low values ($< 2 sigmas$) for small values of $\log r$ it is not possible the rejection of null hypothesis.

### 3.5.2 Correlation dimension of the $V_{2-10}$ SVD component

Fig.10[e-h] is similar to Fig.10[a-d] but for the $V_{2-10}$ SVD component. The slope profiles of this signal reveal clearly low value saturation at value lower $\sim 5.5$ (Fig.10e) with possibility of strong discrimination from the corresponding surrogate data Fig.10[f,g]. The significance of the discriminating statistics is much higher than two sigmas (Fig.10h.). This permits the rejection of the null hypothesis with confidence $> 99\%$.

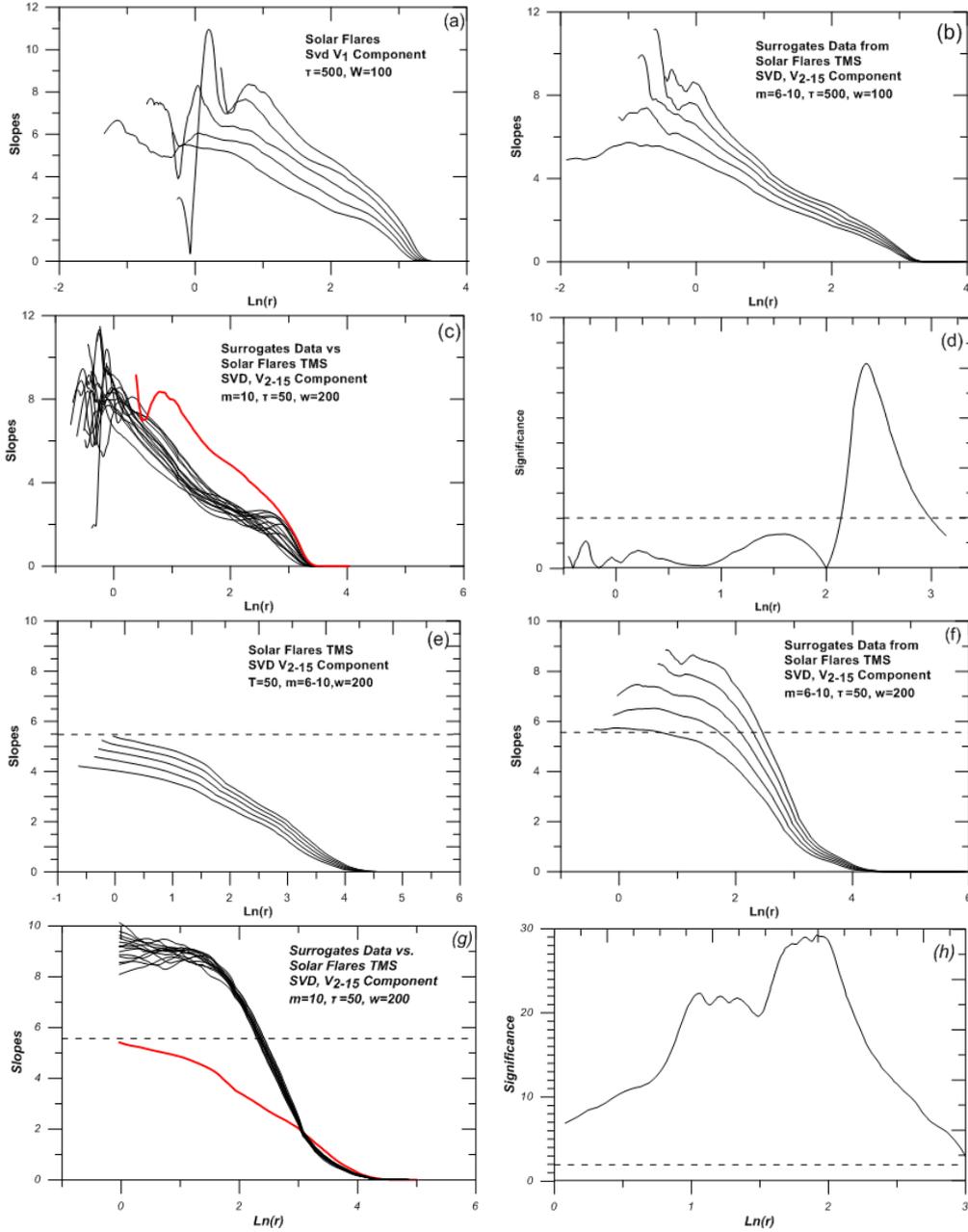

*Figure 10:* (*a*) *Slopes D of the correlation integrals estimated for $V_1$ SVD component.* (*b*) *Slopes D of the correlation integrals estimated for the Surrogate timeseries.* (*c*) *Slopes of the correlation integrals of the $V_1$ SVD component and its thirty (30) surrogate estimated for delay time τ=500 and for embedding dimension m=10, as a function of Ln(r).* (*d*) *The Significance of the Statistics for the $V_1$ SVD component timeseries and its 30 Surrogates.* (*e*) *Slopes D of the correlation integrals estimated for $V_{2-15}$ SVD component.* (*f*) *Slopes D of the correlation integrals estimated for the Surrogate timeseries.* (*g*) *Slopes of the correlation integrals of the $V_{2-15}$ SVD component and its thirty (30) surrogate, estimated for delay time τ=500 and for embedding dimension m=10, as a function of Ln(r).* (*h*) *The Significance of the Statistics for the $V_1$ SVD component timeseries and its 30 Surrogates.*

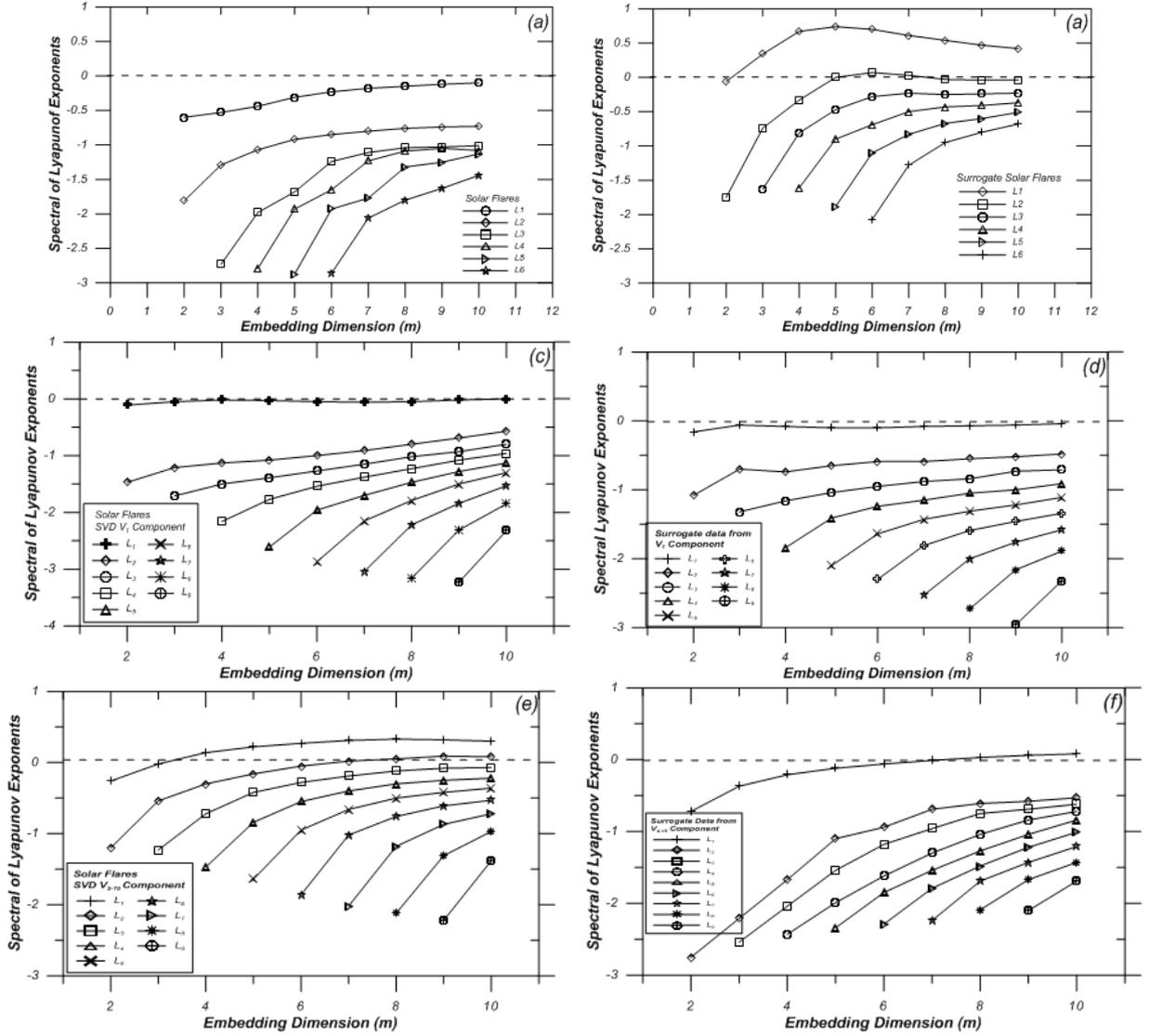

***Figure 11:*** *(a) The spectrum of the Lyapunov exponents $L_i$ as a function of embedding dimension for the Solar Flares timeseries (b) The spectrum of the Lyapunov exponents $L_i$ as a function of embedding dimension for the Surrogate data. (c) The spectrum of the Lyapunov exponents $L_i$ as a function of embedding dimension for the $V_1$ SVD component (d) The spectrum of the Lyapunov exponents $L_i$ as a function of embedding dimension for the Surrogate data. (e) The spectrum of the Lyapunov exponents $L_i$ as a function of embedding dimension for the $V_{2-15}$ SVD component (f) The spectrum of the Lyapunov exponents $L_i$ as a function of embedding dimension for the Surrogate data.*

### 3.6 Determination of the spectra of the Lyapunov exponents

In this section we present the estimated spectra of Lyapunov exponents for the original timeseries of the solar flare index and its SVD components. Fig.11a shows the spectrum $L_i, i = 1-6$ of the Lyapunov exponents estimated for the original timeseries of the solar flares index.
As we can observe in Fig.11a there is no positive Lyapunov exponent, while the largest one approaches the value of zero, from negative values. Fig.11b is similar to Fig.11a but for the surrogate data corresponding to the original timeseries. The similarity of the Lyapunov exponent spectra between the original signal and its surrogate data is obvious. The rejection of the null hypothesis is impossible as the significance of the discriminating statistics was estimated to be lower than two sigmas. Fig.11[c-d] is similar to the previous figures but for the $V_1$ SVD component of the original timeseries. Now, there is strong differentiation from surrogate data, as the significance of the statistics obtains values much higher than two sigmas, while the largest Lyapunov exponent obtains zero value. Finally in Fig.11[e-f], we present the Lyapunov exponent spectrum for the $V_{2-15}$ SVD component (Fig.10[e]) and its surrogate data (Fig.10[f]). In the case of the $V_{2-15}$ SVD component it is also observed the possibility for a strong discrimination from the surrogate data, as the significance of the statistics was found much higher than two sigmas. Here the largest Lyapunov exponent was estimated to be clearly positive.

**Table 1**

|  | Solar Flares TMS | Sunspot Index TMS | Solar Flares V1 Compon. | Sunspot Index V1 Compon. | Solar Flares V2-10 Compon. | Sunspot Index V2-15 Compon. |
|---|---|---|---|---|---|---|
| $\Delta\alpha = \alpha_{max} - \alpha_{min}$ | 2.15 | 1.752 | 0.77 | 1.113 | 1.37 | 1.940 |
| q_sen | 0.308 | 0.368 | -0.540 | 0.055 | 0.192 | 0.407 |
| q_stat | 1.87±0.06 | 1.53 ± 0.04 | 1.28±0.04 | 1.40±0.08 | 2.02±0.15 | 2.12 ± 0.20 |
| q_rel (C(τ)) | 5.329±0.135 | 5.672±0.127 | 42.666±2.950 | 29.571±0.794 | 4.546±0.173 | 4.115±0.134 |
| q_rel (I(τ)) | 9.333±0.395 | 2.522±0.044 | 9.849±0.410 | 5.255±0.308 | 3.207±0.086 | 2.426±0.054 |
| L1 | ≈0 | ≈0 | ≈0 | ≈0 | >0 | >0 |
| Li, (i>2) | <0 | <0 | <0 | <0 | <0 (L2>0) | <0 (L2>0) |
| D (cor. Dim.) | >8-10 | >8 | >6-8 | >6 | ≈5.5 | ≈6 |

*Table 1: Summarize parameter values of solar dynamics including the sunspot and the solar flare dynamics: From the top to the bottom we show: changes of the ranges $\Delta\alpha$ of the multifractal profile. The q-triplet ($q_{sen}, q_{stat}, q_{rel}$) of Tsallis. The values of the maximum Lyapunov exponent ($L_i$), the next Lyapunov exponent and the correlation dimension (D).*

The dynamical characteristics presented in this table indicate clearly discrimination between the sunspot, flare and solar processes concerning their dynamical characteristics.

**4. Summary of Data Analysis Results**

In this study we used the SVD analysis in order to discriminate the dynamical components, underlying the solar flare index timeseries. After this we applied an extended algorithm for the nonlinear analysis of the original solar flare index timeseries, its $V_1$ (first) SVD component and the signal $V_{2-10}$ composed from the sum of the higher SVD components. The analysis was expanded to include the estimation of: a) Flatness coefficients as a measure of Gaussian, non-Gaussian dynamics, b) The $q$–triplet of Tsallis non-extensive statistics, c) The correlation dimension, d) The Lyapunov exponent spectrum, e) The spectrum of the structure function scaling exponent.
The results of data analysis presented in section 3 are summarized as follows:
- Clear distinction was everywhere observed between two dynamics: a) the solar flare dynamics underlying the first ($V_1$) and b) the solar flares dynamics underlying the ($V_{2-10}$) SVD component of the solar flare index timeseries.
- The non Gaussian and non-extensive statistics were found to be effective for the original solar flare index as well as its SVD components $V_1, V_{2-10}$ of the solar flares process, indicating non-extensive solar dynamics.

- The Tsallis $q$–triplet ($q_{sen}, q_{stat}, q_{rel}$) was found at every case to verify the expected scheme $q_{sen} \leq 1 \leq q_{stat} \leq q_{rel}$. Moreover the Tsallis $q$–triplet estimation showed clear distinction between the various dynamics underlying the SVD analyzed solar flare signal as follows: $q_k(V_1) < q_k(original) < q_k(V_{2-10})$, for all the $k \equiv (sen, stat, rel)$.

- The multifractal character was verified for the original signal and its SVD components. Also, the multifractality was found to be intensified as we pass from the $V_1$ to the $V_{2-10}$ SVD component in accordance to the relation: $0 < \Delta a(V_1) < \Delta a(V_{2-10})$, $0 < \Delta Dq(V_1) < \Delta Dq(V_{2-10})$, where $\Delta a = a_{max} - a_{min}$, and $\Delta Dq = D_{q=-\infty} - D_{q=+\infty}$.

- Efficient agreement between $D_q$ and $p$–model was discovered indicating intermittent (multifractal) solar turbulence which is intensified for the solar dynamics underlying the $V_{2-10}$ SVD component of the sunspot index.

- Generally the differences $\Delta a(V_i)$ and $\Delta D_q(V_i)$ increase passing from lower to higher SVD components $V_i, i = 1, 2, ..., 7$.

- The $q_{rel}$ index was estimated for two distinct relaxation magnitudes, the autocorrelation function $C(\tau)$ and the mutual information $I(\tau)$, indicating a good agreement taking into account the linear – nonlinear character of the $C(\tau)$ and $I(\tau)$ respectively.

- The correlation dimension was estimated at low values $D_{CD} = 4 - 5$ for the $V_{2-10}$ SVD component. For the $V_1$ SVD component of the original signal the correlation dimension was found to be higher than the value $\sim 8$.

- Also the null hypothesis of non-chaotic dynamics and non-linear distortion of white-noise was rejected only for the $V_{2-10}$ SVD component. For the original signal and its $V_1$ SVD component the rejection was insignificant ($s < 2$). These results indicate non-linearity low dimensional determinism solar dynamics underlying the $V_{2-10}$ SVD component and high dimensional self organized criticality (SOC) solar dynamics for the $V_1$ SVD component.

- The estimation of Lyapunov exponent spectrum showed for the $V_{2-10}$ SVD component one positive Lyapunov exponent ($\lambda_1 > 0$) clearly discriminated from the signal of surrogate data. For the original sunspot signal and its $V_1$ SVD component the discrimination with surrogate data was inefficient. These results indicate low-dimensional and chaotic deterministic dynamics underlying the $V_{2-10}$ SVD ($\lambda_1 > 0$) component and weak chaos solar dynamics underlying to the $V_1$ SVD component related to a SOC process at the edge of chaos ($\lambda_1 = 0$).

- The structure function scaling exponents spectrum $J(p)$ was estimated to values lower than the corresponding $p/3$ values of the K41 theory, in both cases, the $V_1$ and the $V_{2-10}$ SVD components, as well as for the original signal of the solar flare index. This result indicates the intermittent (multifractal) character of the solar corona turbulence dissipation.

- The slopes $dJ(p)/dp$ of the scaling exponent function were found to be decreasing as the order $p$ increases. This result confirms the intermittent and multifractal character of the low solar corona turbulence dissipation process.
- The SVD analysis of the solar flare dynamics in relation to the structure functions exponent spectrum $J(p)$ estimated for the SVD components showed clearly the distinction of the dynamics underlying the first and the higher SVD components.
- The difference $\Delta J(p)$ between the experimental and theoretical values of the scaling exponent spectrum of the structure functions was found to follow regionally a linear profile: $\Delta J(p) = ap + b$ for the original sunspot index and its $V_1$ and the $V_{2-10}$ SVD components.
- Adding the $\Delta J(p) = ap + b$ function to the theoretically estimated $J(p)$ values by using Tsallis theory we obtain an excellent agreement of the theoretically predicted and the experimentally estimated exponent spectrum values $J(p)$.
- Noticeable agreement of Tsallis theory and the experimental estimation of the functions $f(a)$, $D(q)$, and $J(p)$ was found also.

## 5. Discussion and Theoretical Interpretation

The results of previous data analysis showed clearly:

a) The non-Gaussian and non-extensive statistical character of the low solar corona dynamics underlying the solar flare index time series.
b) The intermittent and multifractal turbulent character of the solar low corona system.
c) The phase transition process between different solar flare dynamical profiles.
d) Novel agreement of $q$ – entropy principle and the experimental estimation of solar flare intermittent turbulence indices: $f(a)$, $D(q)$ and $J(p)$.
e) Clear discrimination of the solar flare dynamics from sunspot dynamics through the q-triplet of Tsallis and structure functions exponent spectrum, as we can conclude by table 1 and Fig.12

The experimental results of this study indicate clearly the non-Gaussianity and non-extensivity, as well as the multi-fractal and multi-scale dynamics of the solar flare process. According to these results we can conclude for solar flares the existence of a new mechanism of anomalous kinetic and magnetic energy dissipation and anomalous charged particle acceleration at the solar flare regions. This mechanism can be characterized as fractal dissipation - fractal acceleration mechanism as the regions of dissipation – acceleration corresponds to fractal fields-particles distributions. This mechanism is further described in the study of Pavlos [8].

It is significant also to notice the similarity between the low solar corona dynamics and the dynamics of the solar convection zone concerning the phase transition process from high dimensional SOC state to low dimensional chaos state. However the discrimination of the two dynamical systems is possible by following the differentiation of the various dynamical characteristics as they are summarized in table 1.

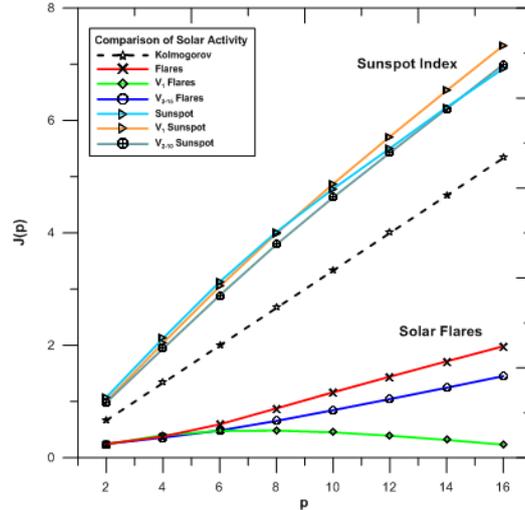

***Figure 12:*** *The J(p) function versus p for sunspot timeseries with $V_1$, $V_{2-10}$ SVD components, Solar Flares timeseries with $V_1$, $V_{2-15}$ SVD components and compared with the Kolmogorov (p/3) prediction (dashed line)*